# Measuring Inaccuracy in Travel Demand Forecasting:

# Methodological Considerations Regarding Ramp Up and Sampling

Bent Flyvbjerg

**All communication to:**

Professor Dr. Bent Flyvbjerg

Department of Development and Planning, Aalborg University

Fibigerstraede 11, 9220 Aalborg, Denmark

Tel: +45 9816 9084 or +45 9635 8375

Fax: +45 9815 3537, Email: flyvbjerg@plan.aau.dk

**Biographical Note**

Bent Flyvbjerg is a professor of planning at Aalborg University, Denmark. He is founding director of the university's research program on large-scale infrastructure projects. He was twice a Visiting Fulbright Scholar to the United States, where he did research at the University of California at Los Angeles and Berkeley and at Harvard University. His latest book is *Megaprojects and Risk* (Cambridge University Press, 2003; with Nils Bruzelius and Werner Rothengatter).

## Abstract

Project promoters, forecasters, and managers sometimes object to two things in measuring inaccuracy in travel demand forecasting: (1) using the forecast made at the time of making the decision to build as the basis for measuring inaccuracy and (2) using traffic during the first year of operations as the basis for measurement. This paper presents the case against both objections. First, if one is interested in learning whether decisions about building transport infrastructure are based on reliable information, then it is exactly the traffic forecasted at the time of making the decision to build that is of interest. Second, although ideally studies should take into account so-called demand "ramp up" over a period of years, the empirical evidence and practical considerations do not support this ideal requirement, at least not for large-N studies. Finally, the paper argues that large samples of inaccuracy in travel demand forecasts are likely to be conservatively biased, i.e., accuracy in travel demand forecasts estimated from such samples would likely be higher than accuracy in travel demand forecasts in the project population. This bias must be taken into account when interpreting the results from statistical analyses of inaccuracy in travel demand forecasting.

## Introduction

As pointed out by Pickrell (1990) and Richmond (1998), estimates of the financial viability of projects are heavily dependent on the accuracy of travel demand forecasts.

Such forecasts are also the basis for socio-economic and environmental appraisal of transportation infrastructure projects. According to the experiences gained with the accuracy of demand forecasting in the transportation sector, covering traffic volumes, spatial traffic distribution, and distribution between transportation modes, there is evidence that demand forecasting--like cost forecasting, and despite all scientific progress in modeling--is a major source of uncertainty and risk in the appraisal of transportation infrastructure projects (Flyvbjerg, Bruzelius, and Rothengatter 2003).

Travel demand forecasts are routinely used to dimension the construction of transportation infrastructure projects. Here accuracy in forecasts is a point of considerable importance to the effective allocation of scarce funds. For example, Bangkok's US$2 billion Skytrain was hugely overdimensioned because the passenger forecast was 2.5 times higher than actual traffic. As a result, station platforms are too long for the shortened trains that now operate the system, a large number of trains and cars are idly parked in the train garage because there is no need for them, terminals are too large, etc. The project company has ended up in financial trouble and even though urban rail is probably a good idea for a congested and air-polluted city like Bangkok, overinvesting in idle capacity is hardly the best way to use resources, and especially not in a developing nation where capital for investment is scarce. Conversely, a UK National Audit Office study identified a number of road projects that were underdimensioned because traffic forecasts were too low. This, too, led to multi-million-dollar inefficiencies, because it is considerably more expensive to add capacity to existing fully used roads than it is to build the capacity up front (National Audit Office 1988). For these and other reasons, accuracy in travel demand forecasts matters, as also pointed out by Hugosson (2005) elsewhere in this issue.

Nevertheless, rigorous studies of accuracy are rare, as are discussions of the methodological problems involved in measuring accuracy. Flyvbjerg, Holm, and Buhl (2005), presents the first statistically significant study of accuracy in travel demand forecasting, based on analyses of 210 forecasts. In what follows, the methodological considerations involved in measuring accuracy are expounded.

## Defining Inaccuracy

In order to estimate the accuracy of travel demand forecasts it is necessary to compare forecasted with actual demand. It is common practice to define the inaccuracy of a demand forecast as actual minus forecasted demand in percentage of forecasted demand:

$$I = ((T_a - T_f) \times 100)/ T_f$$

where

$I$ = inaccuracy in percent

$T_a$ = actual travel demand

$T_f$ = forecasted travel demand

Actual demand is typically counted for the first year of operations (or the opening year). Forecasted demand is typically the estimate for the first year of operations (or the opening year) as estimated at the time of decision to build the project (the time the project was formally given the go-ahead).

Thus the forecast for a given project is the estimate that was available to decision makers when they made the decision to build the project. By measuring inaccuracy in relation to this base year one provides an answer to the question of whether the decision to build was an informed one, that is, whether the decision was based on accurate information, here accurate travel forecasts. If no estimate of travel demand was available at the time of decision to build, then the closest available estimate is typically used, often a later estimate resulting in a conservative bias in the measure for inaccuracy.

With this definition of inaccuracy, perfect accuracy is indicated by zero; an inaccuracy of minus 40 percent, for example, would indicate that actual traffic were 40 percent lower than forecasted traffic, whereas an inaccuracy of plus 40 percent would mean that actual traffic were 40 percent higher than forecasted traffic.

Below, the pros and cons of this way of defining and measuring inaccuracy in travel demand forecasts are discussed.

## Time of Decision to build as Base Year for Measuring Inaccuracy

Planners and promoters of transportation infrastructure projects sometimes object to using the time of decision to build as the base year for measuring inaccuracy (Flyvbjerg, Bruzelius, and Rothengatter 2003). They say various forecasts are made at different stages of planning and implementation with forecasts typically becoming more accurate over time. Thus the forecast at the time of making the decision to build is far from final. It is only to be expected, therefore, that such an early estimate would be highly inaccurate, and it would be unfair to use this estimate as the basis for assessing the accuracy of travel demand forecasting, or so the objection goes.

The objection must be rejected, however, because when the focus is on decision making, and hence on the accuracy of the information available to decision makers, then it is exactly the traffic forecasted at the time of making the decision to build that is of primary interest. Otherwise it would be impossible to evaluate whether decisions are informed or not. Forecasts made after the decision to build are by definition irrelevant to this decision. Whatever the reasons are for inaccurate forecasts, legislators and citizens--or private investors in the case of privately funded projects--are entitled to know the uncertainty of forecasted traffic and revenues. Otherwise transparency, accountability, and good governance will suffer.

It is furthermore observed that if the inaccuracy of early traffic estimates were simply a matter of incomplete information and inherent difficulties in predicting a distant future, as project promoters and forecasters often say it is, then one would expect inaccuracies to be random or close to random. Inaccuracies, however, have a striking non-random bias just as there are remarkable differences in inaccuracy between modes of transportation, as demonstrated by Flyvbjerg, Holm, and Buhl (2005).

## The Problem With Demand Ramp Up

Planners and promoters also sometimes object to using traffic in the first year of operations (or in the opening year) as the basis for measuring inaccuracy in forecasts. A manager at Eurotunnel, the owner and operator of the Channel tunnel, which is one of the projects my colleagues and I have studied, put it in the following manner in a comment on some of our previous work: "[I]t is misleading to make judgements about success or failure based on traffic revenues in the initial start-up years of the project" (letter from Eurotunnel to the author 1999). If projects experience start-up problems, which was very much the case for the Chunnel, then this may initially affect traffic negatively, but it would only be temporarily and it would be misleading to measure inaccuracy of forecasts on that basis, according to this argument. When start-up problems are over, normal operations will ensue, traffic will increase, and this should be the basis on which inaccuracy is measured, the argument continues. Furthermore, it takes time before travelers effectively discover and make use of a new transportation facility and change their travel behavior accordingly. Inertia is a factor. A project with lower-than-forecasted traffic during the first year of operations may well catch up with the forecast a few years down the line and it would be more appropriate to measure inaccuracy on that basis. If the first year of operations is used as the basis for comparison, the result would be the identification of too many underperforming projects, or so the opponents to using this basis argue.

At first sight the argument sounds convincing, and in principle (as opposed to in practice) there is nothing which prevents using another time period than first year of operations as the basis for measuring inaccuracy. One might, for example, decide to use the fifth year of operations, because start-up problems might be expected to be ironed out by then; while important external changes in for instance land use will not have developed fully at this time either.

If demand forecasts do not explicitly model demand for the opening year it is important to be aware of this. Some forecasts assume so-called demand "ramp up" over the first 3-5 years of a scheme up to the full modeled demand. For example 50 percent of total demand in year 1, 80 percent in year 2, etc. By comparing opening years one may thus be assessing two sources of uncertainty: the inaccuracy of the modeling itself, and the inaccuracy of the assumed ramp up to full demand.

In cost-benefit analyses, errors in the ramps up are likely to have a relatively minor impact on the total present value of benefits as compared to errors in the forecast total demand. Therefore the question is whether a focus on the opening year gives too prominent a role to possible ramp-up errors in comparison to their impact on the overall benefit-cost ratios of projects.

Two measures would solve the issue. One would be to use, e.g., the fifth year to compare forecast to actual demand, as mentioned above. The other would be to do statistical tests as to whether there is any difference in bias and variability between the opening year and the fifth year after opening. Such measures are certainly desirable and they may be workable for small-N studies, i.e., studies that cover one or a few projects and where the relevant data have been produced and stored.

For large-N studies, i.e., studies covering many projects, these ideal measures prove unworkable, however, because in practice only few projects can be found for which a travel demand forecast is available for the fifth year of operations *and* actual traffic counts are also available for this year so that inaccuracy may be systematically measured for the year. Available studies with statistical comparisons of the first and the fifth year are even rarer. Many more projects can be found with available information about forecasted and actual traffic for the first year of operations than for later years, because it appears to be the most common practice for both forecasters and those who evaluate the accuracy of forecasts to use first-year-of-operations as the basis for their work (Fouracre et al. 1990, Pickrell 1990, National Audit Office 1992, Walmsley and Pickett 1992, World Bank 1994). For this reason, in large-N studies it is advisable to stay with first year of operations as the basis for measuring inaccuracy. Essentially, the state-of-the-art in this specialized area of transportation research is currently such that large-N studies will be impossible if this advise is ignored. In addition to this line of reasoning the following four arguments support using the first year of operations as the basis for measuring forecasting inaccuracy.

First, for projects for which data are available on actual and forecasted traffic covering more than one year after operations begin, it turns out that projects with lower-than-forecasted traffic during the first year of operations also tend to have lower-than-forecasted traffic in later years. Thus using first year of operations as the basis for measuring inaccuracy appear not often to result in the error of identifying projects as underperforming that would not be identified as such if a different time period were used as the basis for comparison. Actual traffic apparently does not quickly catch up with forecasted traffic for this type of project, and sometimes it never does. Ramp up may initially be assumed, but it does not exist in reality for these projects. For instance, two follow-up studies of the urban rail projects analyzed by Pickrell (1990) showed no significant gains in patronage over time; for Baltimore, Buffalo, and Pittsburgh patronage actually dropped over time (Richmond 1998, Ryan 2004). For the Channel tunnel, more than five years after opening to the public, Eurostar train passengers numbered only 45 percent of that forecasted for the opening year; rail freight traffic was 40 percent of that forecasted; the situation has not improved and the result has been several near-bankruptcies. For the Humber bridge in the UK, 16 years after opening to the public actual traffic was still only about half of that forecasted. In Denmark, it took more than 20 years for actual traffic on the New Little Belt bridge to catch up with forecasted traffic, and for several years the difference between forecasted and actual traffic grew larger instead of smaller. Assuming ramp up for such projects would lead to less accuracy in demand modeling, not more. These findings fit well with Mierzejewski's (1995: 31-32) observation that the conventional wisdom in forecasting, that in the long run forecasting errors tend to cancel each other out, is wrong; errors often reinforce each other with the result that inaccuracy becomes larger when measured against later years as compared to when measured against the first year of operations. Following this logic, using first year of operations as the basis for measuring inaccuracy would tend to underestimate overall inaccuracy of travel demand forecasts.

Second, sightseeing traffic may be substantial during the first months of operations for the type of large-scale transportation infrastructure project my colleagues and I have focused on in our studies, many of which are architectural and engineering marvels, in addition to being prosaic transportation machines designed to efficiently get people and goods from point A to point B. Sightseeing traffic is traffic attracted by a project on the basis of people's desire to see and try the new transportation facility in question, for instance a new bridge or a new rail line. To illustrate, for the Øresund bridge between Sweden and Denmark, road traffic during the first month of operations was 19 percent higher than traffic for the same month one year later. The difference between the two months can mainly be ascribed to sightseeing traffic, which was somewhat lower than expected by the project company (Trafikministeriet,

Finansministeriet, and Sund & Bælt Holding, Ltd. 2002: App. 4:2). Sightseeing traffic may help offset the possible negative impacts on travel demand from start-up problems etc. mentioned above, at least for projects that are sufficiently attractive in the public's eye to warrant sightseeing. The existence of such countervailing influences on traffic during the start-up phase of projects help explain why first-year-of-operations tend to be a fairly precise basis for measuring inaccuracy in travel demand forecasts for many projects.

Third, it may be observed as an empirical fact that forecasters and planners typically use first-year-of-operations as the principal basis for making their forecasts. For a given project, this is generally the main forecast presented to decision makers and it forms part of the information decision makers have at hand in making their decision of whether to build or not. If one wants to evaluate whether such decisions are informed, then it is the accuracy of this forecast that must be evaluated and one must therefore compare actual traffic in the first year of operations with forecasted traffic for that year.

Fourth and finally, if newly opened transport infrastructure projects have a systematic ramp up period before traffic picks up, as claimed by many planners and promoters, this should be empirically documented and integrated in travel demand modeling. In this way adaption would be reflected in forecasts instead of being external to these.

In sum, project promoters, forecasters, and managers sometimes object to two things in measuring inaccuracy of travel demand forecasts: (1) using the travel demand forecast made at the time of making the decision to build as the basis for measuring inaccuracy and (2) using traffic during the first year of operations as the basis for measurement. However, if one were to follow the objections then it would be virtually impossible to make meaningful comparisons of forecasted and actual traffic across large numbers of projects and across project types, geographical areas, and time periods because no common standard of comparison would be available. Finally, the methodology proposed here as a common standard is already widely used in practice for measuring the inaccuracy of travel demand forecasts. This method conveniently allows meaningful and consistent comparisons across space and time.

## Issues of Sampling, Data Collection, and Bias

Data that allow the calculation of inaccuracies in travel demand forecasts unfortunately are relatively rare. For public sector projects, often the data are simply not produced. Even where the intention is to produce the data, projects may develop in ways that make it difficult or impossible to compare forecasted with actual traffic. For example, forecasted traffic for a given project may be estimated for the opening year, but due to delays, which are common, the actual opening date turns out to be several years later than that forecasted, and no forecast of traffic was made for the actual opening year. In more general terms, methodological differences in how and on what basis forecasted and actual traffic are estimated often make comparisons difficult or impossible. Finally, for large projects the elapse of time from forecasts are made, until decision to build, until construction starts, until the project is completed, until operations begin, and until actual traffic can finally be counted may cover five, ten, or more years. Over such long time periods the assumptions underlying forecasts may be dated and incommensurate when compared to the assumptions underlying the way actual traffic is measured, or initial plans to compare actual with forecasted traffic may be given up or simply forgotten.

For private sector projects, traffic typically generates an income for the project owner. Budgeting and accounting is commercial, and therefore travel demand forecasts and traffic counts tend to be more systematic and more conducive to comparative studies of forecasted and actual traffic than is the case for public sector projects. This typically does not help scholars much, however, because traffic data in private projects are often classified to keep them from the hands of competitors. And for both public and private projects, data that allow forecasted and actual traffic to be compared may be held back by project owners and managers because the size and direction of differences between forecasted and actual traffic may be of a kind that, if made public, would make the project look bad in the public eye, for instance where actual traffic is substantially lower than that forecasted, which is a common situation.

Because of this type of problem with data availability and access, scholarly studies of the accuracy of travel demand forecasts are rare. Where such studies exist, they are characteristically small-N research, i.e., they are single case studies or they cover only a small sample of projects (Fouracre et al. 1990, Pickrell 1990, National Audit Office 1992, Walmsley and Pickett 1992, World Bank 1994). Consequently, with existing data and studies, it has so far been impossible to give statistically satisfying answers to questions about how accurate travel demand forecasts are for transportation infrastructure projects. As a consequence, planners and analysts have been unable to deliver valid and reliable risk assessments for travel demand, even though such assessments are a requirement for informed financial appraisal of transportation infrastructure projects.

As mentioned previously, the study of demand forecasts in 210 transportation infrastructure projects reported in Flyvbjerg, Holm, and Buhl (2005) is the first survey sufficiently large to permit statistically valid answers to questions regarding the accuracy of travel demand forecasts. To establish a relatively large sample like this, because data are so scarce their availability becomes a main criterion for data collection. But just because data are available this does not mean they are valid, reliable, and comparable, as they must be for systematic analysis. To illustrate, the sample of forecasts in the 210 projects was established on the basis of data from a much larger group of projects, the majority of which were rejected because of unclear or insufficient data quality. Of 485 projects considered, 275 were rejected in this manner. Of the 275 rejects, 151 projects were rejected because inaccuracies for travel demand forecasts in these projects turned out to have been estimated on the basis of adjusted data for actual traffic instead of using original, real count data, which is more valid; 124 projects were rejected because inaccuracy had been estimated in ways different from and incomparable to the way inaccuracy is defined above. In other words, of the 485 observations only 210 were considered valid, reliable, and comparable. Such quality control of data and rejection of data with low quality is crucial to arriving at useful results when measuring inaccuracy in travel demand forecasting.

For any sample, a key question is whether it is representative of the population. For a sample where data availability is crucial to sampling, this question translates into one of whether projects with available data are likely to be representative. There are four reasons why this question must be answered in the negative for transportation infrastructure projects with data on the inaccuracy of demand forecasts.

First, it has been argued that the very existence of data that make the evaluation of performance possible--here performance regarding the accuracy of forecasts--may contribute to improved performance when such data are used by project management to monitor projects (World Bank 1994: 17). Such projects would have better than average, i.e. non-representative, performance.

Second, one might speculate that managers and promoters of projects with a particularly bad track record regarding travel demand forecasts have an interest in not making traffic data available, which would then result in underrepresentation of such projects in the sample. Conversely, managers and promoters of projects with a good track record for travel demand forecasts might be interested in making this public, resulting in overrepresentation of these projects.

Third, even where managers have made traffic data available they may have chosen to give out data that present their projects in as favorable a light as possible. Often there are several forecasts of traffic to choose from and several counts or compilations of actual traffic for a given project at a given time. If researchers collect data by means of survey questionnaires, as is often the case, there might be a temptation for managers to choose the combination of forecasted and actual traffic that suits them best, possibly a combination that makes their projects look good. An experienced researcher in a large European country, who was giving us feedback on our research for that country, commented on the data collection (the quote has been anonymized for obvious reasons):

> "Most of the [research] is based on [national railway] replies to a questionnaire. This is likely to create a systematic bias. [The national railways] cannot be trusted to tell you the truth on these matters. As you know very well, the concept of 'truth' in these matters is particularly fragile. The temptation for [the national railways] to take, for the forecasts, the number that suits them best, this temptation must be great, and I don't think they could resist it. What you would need [in order to obtain better data] would be the original forecast documents, preferably from the archives of the Ministry of Transportation (not [from the national railways]), that were utilised to take the decision."

Other studies have documented the existence of such "cooking" of data (Wachs 1989, 1990). Unfortunately, in practice it proves difficult and often impossible to gain access to the original forecast documents. This is why researchers sometimes have to rely on the second-best methodology of survey questionnaires. This is also why available data may not be representative data.

Fourth, and finally, differences in the representativity of different subsamples may also result in non-representative data, for instance differences between rail and road.

The available data do not allow an exact, empirical assessment of the magnitude of the problem of misrepresentation. It must be concluded, however, for the reasons given above, that given the current state of the art large samples of forecasted and actual traffic are likely to be biased and the bias is likely to be conservative. In other words, accuracy in travel demand forecasts estimated from such samples would likely be higher than accuracy in travel demand forecasts in the project population. This must be kept in mind when interpreting the results from statistical analyses of inaccuracy in travel demand forecasting.